\RequirePackage[final]{graphicx}
\documentclass[10pt, conference, draft]{IEEEtran}

\IEEEoverridecommandlockouts

\usepackage{ifdraft}
\usepackage[export]{adjustbox}
\usepackage[T1]{fontenc}
\usepackage{fixltx2e}
\usepackage{amsmath}
\usepackage{amssymb}
\usepackage{mathrsfs}
\usepackage{hyperref}
\usepackage{pxfonts,wasysym}
\usepackage{yfonts}
\usepackage[plain]{fancyref}
\usepackage{subcaption}

\usepackage[final]{listings}
\usepackage{booktabs}
\usepackage{threeparttable}
\usepackage[usenames, dvipsnames]{xcolor,colortbl}
\usepackage{xspace}
\usepackage{etoolbox}
\usepackage[inline]{enumitem}
\usepackage{acronym}
\usepackage{url}
\usepackage{xcolor}
\usepackage{longtable}
\usepackage{multirow}

\usepackage[backend=biber,
   style=ieee,
   natbib=true,
   maxcitenames=1,
   mincitenames=1,
   hyperref=true,
   url=false,
   doi=false,
   sorting=nyt,
   dashed=false,
   defernumbers]{biblatex}

\AtEveryBibitem
{
   \clearlist{address}
   \clearfield{date}
   \clearfield{eprint}
   \clearfield{issn}
   \clearfield{month}
   \clearfield{note}
   \clearfield{arxivId}
   \clearlist{location}
   \clearfield{number}
   \clearfield{series}
   \clearfield{url}
   \clearname{editor}
   \ifentrytype{inproceedings}
     {\clearfield{day}
      \clearfield{month}
      \clearfield{volume}}{}
}

\DeclareFieldFormat*{title}{\textsl{#1}\isdot}
\DeclareFieldFormat*{journaltitle}{#1}
\DeclareFieldFormat*{booktitle}{#1}

\renewbibmacro{in:}{} 
\renewbibmacro{ser.}{} 

\DeclareSourcemap
 {\maps[datatype=bibtex,overwrite]
   {
    \map
      {\step[fieldsource=publisher,
       match=\regexp{Association for Computing Machinery}, replace={ACM}]}
    \map
      {\step[fieldsource=booktitle,
       match=\regexp{[Pp]roceedings}, replace={Proc.}]}
    \map
      {\step[fieldsource=booktitle,
       match=\regexp{[Ii]nternational}, replace={Intl.}]}
   \map
      {\step[fieldsource=booktitle,
       match=\regexp{[Cc]onference}, replace={Conf.}]}
    \map
      {\step[fieldsource=booktitle,
       match=\regexp{[Ss]ymposium}, replace={Symp.}]}
    \map
      {\step[fieldsource=journal,
       match=\regexp{[Jj]ournal}, replace={Jour.}]}
    \map
      {\step[fieldsource=journal,
       match=\regexp{[Ii]nternational}, replace={Intl.}]}
    \map
      {\step[fieldsource=journal,
       match=\regexp{[Tt]ransactions}, replace={Trans.}]}
    \map
      {\step[fieldsource=journal,
       match=\regexp{[Aa]utonomous and Adaptive Systems}, replace={Auton. Adapt. Syst.}]}
    \map
      {\step[fieldsource=booktitle,
       match=\regexp{[Pp]roceedings\s+of\s+the.+[Ee]uropean\s+[Cc]onference\s+in}, replace={European Conf. in}]}
    \map
      {\step[fieldsource=booktitle,
       match=\regexp{In\s+[Pp]roceedings\s+of\s+the\s+[Ss]ymposium\s+on}, replace={Symp. on}]}
     \map
      {\step[fieldsource=publisher,
       match=\regexp{[Aa]ssociation\s+for\s+[Cc]omputing\s+[Mm]achinery\s}, replace={ACM}]}
     \map
      {\step[fieldsource=booktitle,
       match=\regexp{[Pp]roceedings\s+of\s+the\s+[Ii]nternational\s+[Cc]onference\s+on}, replace={Intl. Conf. on}]}
    \map
      {\step[fieldsource=booktitle,
       match=\regexp{[Pp]roceedings\s+of\s+the\s+[Ii]nternational\s+[Ww]orkshop\s+on}, replace={Intl. Workshop on}]}}}

\newtheorem{ex}{Example}

\let\origEx\ex
\let\origendEx\endex
\renewenvironment{ex}{\origEx\upshape}{\origendEx}

\definecolor{author}{rgb}{.0, .0, .0}
\definecolor{comment}{rgb}{.9, .1, .0}
\definecolor{note}{rgb}{.2, .4, .7}
\definecolor{idea}{rgb}{.1, .7, .0}
\definecolor{OliveGreen}{rgb}{0,0.6,0.3}

\def\fref{\Fref} 
\renewcommand{\lstlistingname}{Snippet}

\newcommand*{\fancyreflstlabelprefix}{lst} 
\newcommand*{\Freflstname}{\lstlistingname}
\newcommand*{\freflstname}{\lstlistingname}
\Frefformat{vario}{\fancyreflstlabelprefix}%
  {\Freflstname\fancyrefdefaultspacing#1#3}
\frefformat{vario}{\fancyreflstlabelprefix}%
  {\freflstname\fancyrefdefaultspacing#1#3}
\Frefformat{plain}{\fancyreflstlabelprefix}%
  {\Freflstname\fancyrefdefaultspacing#1}
\frefformat{plain}{\fancyreflstlabelprefix}%
  {\freflstname\fancyrefdefaultspacing#1}

\newcommand*{\fancyrefthmlabelprefix}{thm} 
\newcommand*{\Frefthmname}{Theorem}%
\newcommand*{\frefthmname}{%
 \MakeLowercase{\Frefthmname}}%
\Frefformat{vario}{\fancyrefthmlabelprefix}%
  {\Frefthmname\fancyrefdefaultspacing#1#3}
\frefformat{vario}{\fancyrefthmlabelprefix}%
  {\frefthmname\fancyrefdefaultspacing#1#3}
\Frefformat{plain}{\fancyrefthmlabelprefix}%
  {\Frefthmname\fancyrefdefaultspacing#1}
\frefformat{plain}{\fancyrefthmlabelprefix}%
  {\frefthmname\fancyrefdefaultspacing#1}

\newcommand*{\fancyreflemlabelprefix}{lem} 
\newcommand*{\Freflemname}{Lemma}%
\newcommand*{\freflemname}{%
 \MakeLowercase{\Freflemname}}%
\Frefformat{vario}{\fancyreflemlabelprefix}%
  {\Freflemname\fancyrefdefaultspacing#1#3}
\frefformat{vario}{\fancyreflemlabelprefix}%
  {\freflemname\fancyrefdefaultspacing#1#3}
\Frefformat{plain}{\fancyreflemlabelprefix}%
  {\Freflemname\fancyrefdefaultspacing#1}
\frefformat{plain}{\fancyreflemlabelprefix}%
  {\freflemname\fancyrefdefaultspacing#1}

\newcommand*{\fancyrefdeflabelprefix}{def} 
\newcommand*{\Frefdefname}{Definition}%
\newcommand*{\frefdefname}{%
 \MakeLowercase{\Frefdefname}}%
\Frefformat{vario}{\fancyrefdeflabelprefix}%
  {\Frefdefname\fancyrefdefaultspacing#1#3}
\frefformat{vario}{\fancyrefdeflabelprefix}%
  {\frefdefname\fancyrefdefaultspacing#1#3}
\Frefformat{plain}{\fancyrefdeflabelprefix}%
  {\Frefdefname\fancyrefdefaultspacing#1}
\frefformat{plain}{\fancyrefdeflabelprefix}%
  {\frefdefname\fancyrefdefaultspacing#1}

\newcommand*{\fancyreflnlabelprefix}{ln} 
\newcommand*{\Freflnname}{Line}%
\newcommand*{\freflnname}{%
 \MakeLowercase{\Freflnname}}%
\Frefformat{vario}{\fancyreflnlabelprefix}%
  {\Freflnname\fancyrefdefaultspacing#1#3}
\frefformat{vario}{\fancyreflnlabelprefix}%
  {\freflnname\fancyrefdefaultspacing#1#3}
\Frefformat{plain}{\fancyreflnlabelprefix}%
  {\Freflnname\fancyrefdefaultspacing#1}
\frefformat{plain}{\fancyreflnlabelprefix}%
  {\freflnname\fancyrefdefaultspacing#1}

\lstdefinestyle{floating}
 {frame=lines,
  float=htb,
  captionpos=b,
  abovecaptionskip=-0pt}

\lstdefinelanguage{Scala} {   
	numbers=left,
	breaklines=true,
	extendedchars=true,
	tabsize=2,
	columns=fullflexible,
	showtabs=false,
	showstringspaces=false,
	showspaces=false,
	showstringspaces=false,
	identifierstyle={\ttfamily},
	keywordstyle={\color{blue}},
	ndkeywordstyle={\color{blue}},
	stringstyle={\color{red}},
	ndkeywords={Int, Bool, String},
	keywords={ abstract,case,catch,class,def,do,else,extends,%
          false,final,finally,for,forSome,if,implicit,import,lazy,%
          match,new,null,object,override,package,private,protected,%
          return,sealed,super,this,throw,trait,true,try,type,%
          val,var,while,with,yield,SetActiveVariations,observe},
	otherkeywords={=>,<-,<\%,<:,>:,\#,@,!},
  sensitive=true,
  morecomment=[l]{//},
  morecomment=[n]{/*}{*/},
}

\lstdefinestyle{scala}
 {language=Scala,
  showstringspaces=false,
  tabsize=2,
  style=floating,
  morekeywords={}
}

\lstnewenvironment{scala}[1][]
 {\lstset{style=scala,#1}}{}  

\newcommand{\scode}[1]{\lstinline[style=scala]{#1}}

\usepackage[version=0.96]{pgf}
\usepackage{tikz}
\usetikzlibrary{arrows, shapes, backgrounds}
\usetikzlibrary{decorations.pathreplacing}
\usetikzlibrary{shapes.misc}
\usetikzlibrary{petri}\tikzstyle{place}=[circle,thick,draw=black!75,minimum size=5mm]
\tikzstyle{iplace}=[circle,dashed,thick,draw=black!75,minimum size=5mm]
\tikzstyle{itransition}=[rectangle,draw,thick,fill=black,minimum size=1mm]
\tikzstyle{etransition}=[rectangle,draw,thick,minimum size=1mm]
\tikzstyle{ctransition}=[rectangle,draw,color=black!45,thick,fill=black!45,minimum size=1mm]

\tikzstyle{copn}=
 [node distance=1.3cm, >=stealth', bend angle=45, auto,
  font=\fontsize{8}{8}\selectfont]

\usetikzlibrary{positioning,shadows,trees,mindmap}
\usepackage[edges]{forest}
\usetikzlibrary{arrows.meta}
\colorlet{linecol}{black!75}
\usepackage{xkcdcolors} 
\usetikzlibrary{tikzmark}
\usetikzlibrary{calc}
\newcommand{\highlight}[2]{\colorbox{#1!17}{$\displaystyle #2$}}

\renewcommand{\highlight}[2]{\colorbox{#1!17}{#2}}

\newcommand{\commentedout}[1]{}

\newcounter{challenge}
\newcommand{\eg}{\emph{e.g.,}\xspace}
\newcommand{\ie}{\emph{i.e.,}\xspace}

\newcommand{\deltaiot}{DeltaIoT\xspace}

\newcommand{\copn}[7][]
  {\ensuremath{\mathopen{<}
      {#2}_{#1},{#3}_{#1},{#4}_{#1},{#5}_{#1},{#6}_{#1},{#7}_{#1}
      \mathclose{>}}}

\WithSuffix\newcommand\copn*[1][]{\copn[#1]{P}{T}{f}{f_\circ}{\rho}{m_0}}

\definecolor{author}{rgb}{.5, .5, .5}
\definecolor{comment}{rgb}{.1, .0, .9}
\definecolor{note}{rgb}{.9, .4, .0}
\definecolor{idea}{rgb}{.1, .7, .0}
\definecolor{missing}{rgb}{.9, .1, .0}

\makeatother

\acrodef{API}{Application Programming Interface}
\acrodef{AOP}{Aspect-oriented Programming}
\acrodef{CAS}{Collective Adaptive Systems}
\acrodef{COP}{Context-oriented Programming}
\acrodef{COPN}[CoPN]{Context Petri net}
\acrodef{DCOPN}[DCoPN]{Distributed Context Petri net}
\acrodef{CPU}{Central Processing Unit}
\acrodef{DSL}{Domain-Specific Language}
\acrodef{FOP}{Feature-Oriented Programming}
\acrodef{FP}{Functional programming}
\acrodef{IOT}[IoT]{Internet of Things}
\acrodef{MAPE}{Monitor, Analyze, Plan, and Execute}
\acrodef{OOP}{Object-Oriented Programming}
\acrodef{RL}{Reinforcement Learning}
\acrodef{SPLE}{Software Product Line Engineering}
\acrodef{POI}[POIs]{Points of Interest}
\acrodef{SAT}{boolean SATisfiability problem}
\acrodef{CIA}[ComInA]{Composing Interacting Adaptations}

\newcommand{\acResetNonTrivial}
  {\acresetall
   \acused{CPU}
   \acused{API}
   \acused{LAN}
   \acused{SMT}
   \acused{GUI}}

\acResetNonTrivial

\begin{document}

\title{Learning Recovery Strategies for Dynamic Self-healing in Reactive Systems}

\author{\IEEEauthorblockN{Mateo Sanabria,$^1$ Ivana Dusparic,$^2$ Nicol\'as Cardozo$^1$}
\IEEEauthorblockA{$^1$Systems and Computing Engineering Department, 
Universidad de los Andes, Colombia \\
$^2$School of Computer Science and Statistics, Trinity College Dublin, Ireland \\
$^1$\{m.sanabriaa,n.cardozo\}@uniandes.edu.co, $^2$ivana.dusparic@tcd.ie} 
}

\maketitle

\begin{abstract}
Self-healing systems depend on following a set of predefined instructions to 
recover from a known failure state. Failure states are generally detected based on domain specific 
specialized metrics. Failure fixes are applied at predefined application 
hooks that are not sufficiently expressive to manage different failure types. Self-healing is usually 
applied in the context of distributed systems, where the detection of failures is constrained 
to communication problems, and resolution strategies often consist of replacing complete 
components.
However, current complex systems may reach failure states at a fine granularity not anticipated by 
developers (for example, value range changes for data streaming in IoT systems), making them 
unsuitable for existing self-healing techniques. 
To counter these problems, in this paper we propose a new self-healing framework that learns recovery 
strategies for healing fine-grained system behavior at run time.
Our proposal targets complex reactive systems, defining monitors as predicates specifying 
satisfiability conditions of system properties. Such monitors 
are functionally expressive and can be defined at run time to detect failure states at any 
execution point. Once failure states are detected, we use a Reinforcement Learning-based 
technique to learn a recovery strategy based on users' corrective sequences. Finally, to execute 
the learned strategies, we extract them as \acl{COP} variations that activate dynamically 
whenever the failure state is detected, overwriting the base system behavior with the recovery 
strategy for that state. We validate the feasibility and effectiveness of our framework through a 
prototypical reactive application for tracking mouse movements, and the DeltaIoT exemplar for 
self-healing systems. Our results demonstrate that with just the definition of monitors, the system is 
effective in detecting and recovering from failures between $55\%-92\%$ of the cases in the first 
application, and at par with the predefined strategies in the second application.
\end{abstract}

\begin{IEEEkeywords}
Self-healing systems, 
Context-oriented Programming, 
Functional-reactive programming, 
RL
\end{IEEEkeywords}

\IEEEpeerreviewmaketitle


\section{Introduction}
\label{sec:introduction}

Self-healing properties enable systems to autonomously recover from failure states. 
To enable self-healing in a software system there are four challenges~\cite{rodosek2009self} to 
address:
\begin{enumerate*}[label=(\arabic*)]
\item build the means to \emph{monitor} the system state,
\item use \emph{fault analysis} to detect the root cause of faults based on the monitored state, 
\item \emph{make a decision} about the detected fault (\ie decide whether the fault is blocking), and
\item execute the \emph{recovery} actions to take the system back to a correct state.
\end{enumerate*}
To execute the aforementioned process, software systems must be 
equipped with the tools and capabilities for each of the tasks. To 
capture the system state and to detect faults therein, self-healing 
systems specify hooks in the system as control points to observe the 
system state. Such hooks are also used to plug the corrective actions 
(\eg alternative execution paths, introductions of new/fixed 
modules)~\cite{rodosek2009self}.
Such characteristics of self-healing systems are problematic, as to diagnose and heal from failures, 
we require them to be anticipated by developers at design time. This implies that self-healing 
systems require a pre-defined set of instructions to recover from prescribe failures at defined points 
in the system's execution. This characteristic prescribes a low flexibility of self-healing systems,  
offering a solution only for known unknowns~\cite{weyns17}. 
Unknown unknowns are not covered by self-healing, leading to system failures.  
Moreover, it is a concern that the resolution of local changes cannot assure global 
healing~\cite{10.1145/361179.361202}, hindering the applicability of self-healing systems. 

We note that self-healing systems have been applied mostly in distributed systems, taking 
advantage of the characteristics of such systems with defined metrics that point 
to possible failure states during the execution. However, the use of 
healing strategies in other systems is still lacking~\cite{Psaier2010,schneider2015survey}.  
In line with this observation, we recognize there is a wide variety of 
research and application opportunities of self-healing systems. For 
example, data streaming and big data~\cite{dundar16}, complex reconfigurable network 
environments~\cite{DBLP:journals/corr/abs-1305-4675}, ensuring service quality in web 
services~\cite{naccache2007self}, or avoid faults requiring an operating system to
restart~\cite{david2007building}.

This paper proposes an alternative to deal with the 
challenges of defining and designing self-healing systems. Our proposal 
breaks the assumptions previously made for monitoring, fault analysis, 
and recovery action execution, about the pre-definition of healing 
strategies (\fref{sec:self-healing}). 
To do this, we first introduce flexible dynamic monitors, that can be 
defined and modified at run time. Moreover, monitors express different conditions as predicates, 
evaluating any system property, which avoids fixed monitoring points. Second, upon detecting 
failures, we use an algorithm based on \ac{RL} options~\cite{randlov98} to learn the self-healing strategies based on
corrective actions taken by a human actor, in order to restore 
the system from the failing state. Once a resolution strategy is 
learned, using \ac{COP}~\cite{salvaneschi+12survey}, the system 
dynamically adapts the execution path whenever the state leading to 
the error is detected again by the monitor. The adapted behavior corresponds 
to the learned recovery strategy.  
Furthermore, our proposal seeks to broaden the application of 
self-healing systems to a new domain, reactive systems~\cite{wan00}. 
The objective behind this, is to evaluate the 
feasibility to apply the process of self-healing systems in a domain 
in which systems are in continuous execution, and in which there is 
not a pre-defined set of metrics to assess the system performance 
across multiple systems. 

Our proposal is realized on top of three main concepts to 
achieve run-time learning and definition of recovery strategies 
(\fref{sec:background}):
\begin{enumerate*}[label=(\arabic*)]
\item reactive systems, 
\item \ac{RL} options, and 
\item \ac{COP}.
\end{enumerate*}

To illustrate the idea of learned self-healing strategies,
consider the running example of a mouse move tracking 
reactive application. In this application we track the mouse position $(x,y)$
as it moves on a $100 \times 100$ GUI, displaying an ASCII value 
based on the $(x,y)$ position.
In the example, we define mouse prohibited areas in the GUI by tessellation 
predicates. Whenever the mouse hovers over 
these areas, it is considered as a fault state.

\begin{figure}[htpb]
  \centering
  \begin{subfigure}{0.4\columnwidth}
    \includegraphics[width=\textwidth]{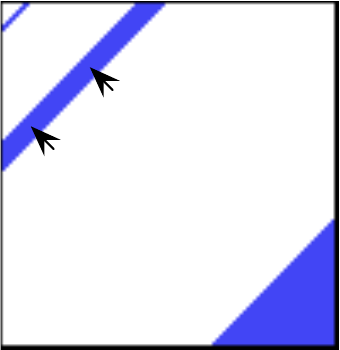}
    \caption{Mouse on a fault state.}
    \label{fig:error}
\end{subfigure}
\hfill
\begin{subfigure}{0.4\columnwidth}
    \includegraphics[width=\textwidth]{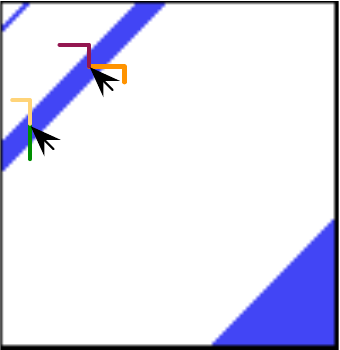}
    \caption{Possible recovery strategy movements.}
    \label{fig:recovery}
\end{subfigure}
  \caption{Mouse position tracking running example}
  \label{fig:running}
\end{figure}

Given that we always want to display the ASCII code of the mouse's 
position, we want to leave the fault state as soon as possible. 
Therefore, a recovery strategy of this behavior is to move the mouse 
away from the fault areas, to a correct execution area. In the example, we assume that the 
the fault areas are unknown beforehand, in consequence, it is not possible to define a recovery 
action for each possible position in the GUI. For each point 
in the fault areas we must define mouse movements to go to a safe 
area. \fref{fig:running} shows the setting of our running example. In 
\fref{fig:error} we show two possible situations in which the mouse is 
in a fault area, \fref{fig:recovery} shows possible resolution 
strategies as paths from the fault states to correct states. The 
better resolution strategy depends on the specific application, 
however, knowing this information beforehand, may not always 
be possible. Our approach lets us learn the best option at run time.

We validate the feasibility and effectiveness of our approach by means of  
two case studies (\fref{sec:validation}). First we use our
running example taken from the reactive systems literature.  
For this application, we validate the feasibility of our solution by introducing monitors 
to detect fault areas. The healing strategy 
is to move the mouse out of the fault area. The results of 
the evaluation show that, using our framework, we are able to detect on average
over 70\% of all failure states, effectively generating and learning a 
healing strategy for each of them. Our second case study reuses the 
DeltaIoT exemplar~\cite{iftikhar17} that dynamically adapts an \ac{IOT} network configuration to heal 
from errors. We use this case study to validate the effectiveness of our proposal to find appropriate 
solutions. We equip the servers in the exemplar with reactive 
characteristics to define predicate monitors to learn healing 
strategies. Our results demonstrate that indeed, the 
learned healing strategies correspond to the adaptations originally proposed in the 
system for the pre-conceived situations. Moreover, we are able to heal 
from situations previously unknown, for which the original system 
cannot recover, demonstrating the effectiveness of our work.

The main contributions of our work are:
\begin{itemize}
  \item Introduction of flexible monitors to detect failures at 
  any moment in time, and any point in the execution, 
  rather than at specified program hooks.
  \item Learning of recovery strategies from failure states, realized 
  as modular context-dependent adaptations, rather than as predefined behavior. 
  \item Application of self-healing techniques to the domain of reactive systems.
\end{itemize}


\section{Background}
\label{sec:background}

To learn self-healing sequences in reactive systems we follow the 
ideas of adaptation generation from sequences of known behavior as achieved by a technique 
called Auto-\ac{COP}~\cite{2103-06757,cardozo22jot}. In this section, we provide the background 
necessary for presenting our solution, specifically Reactive Systems and the two main concepts that 
Auto-\ac{COP} encompasses: \ac{COP} and \ac{RL} options.

\subsection{Reactive Systems}
\label{sec:REScala}

Reactive systems are based on the prompt reaction to internal (\ie system) and external (\ie in the 
surrounding system environment) events. There are two main approaches to address the 
complexity of reactive applications: event-based languages, and 
languages with direct representation of reactive values. Hereinafter we focus on the second approach.

Reactive values (or \emph{behaviors})~\cite{ElliottHudak97:Fran} are introduced in reactive 
programming languages to abstract values that 
vary continuously over time. \textit{Behaviors} can also be defined as dynamic dependent objects 
on other time-varying entities. In this case, \textit{behaviors} update in response to updates from  
\textit{behaviors} they depend on. 
The added value of reactive programming is that now programmers can avoid explicit value updates; 
instead, values are updated automatically, as the language runtime takes care of the 
complexity of updating values asynchronously. 

REScala~\cite{salvaneschi2014rescala} is a reactive 
language built on top of Scala, that provides a robust event system with 
seamless integration support for reactive values, promoting a mixture 
of \ac{OOP} and \ac{FP}. REScala supports \emph{Signals} to
express functional dependencies among values in a declarative way, 
and \emph{events} for continuous or discrete-time changing values. A Signal 
represents the state in the application, whereas events hold  
values that change when fired, representing actions in the 
application. REScala exposes a series of abstractions for signals and 
events, uniformly applying over them.

To propagate changes through a reactive system, REScala provides 
observers which attach a handler function to the event. Every time an 
event fires, its handler function is applied to the current event 
value.

The following snippet shows an observer example. Line 1 
associates the signal \scode{counting} with a handler function.  
\scode{counting} is an integer signal that holds the monitor's 
learning steps so far. Note that the handler function (Lines 2-3) 
changes the value of the learning flag variable when the learning 
steps limit is reached.  

\begin{lstlisting}[language=Scala, numbers=left, label={lst:observerExample}]
   val learningStageObserver = counting observe 
   (count => { if (count > learningStepsLimit) 
        learningFlag = false})
\end{lstlisting}

Observers are the base for failure detection inside monitors in our 
framework. As we explain in \fref{sec:error-detection}, the handle 
function corresponds to user-defined predicates that specify  
failure states. Whenever the predicate evaluates to \scode{true}, the monitor 
triggers the corresponding learned healing strategy.

\subsection{Context-Oriented Programming}
\label{sec:ContextScala}

\ac{COP} is a programming paradigm which enables modeling of the variability required by complex adaptive systems based on behavioral variations that depend on the surrounding 
execution context~\cite{hirschfeld2008context}. Behavioral variations 
are defined by \textit{layers} associated with 
partial method definitions. \textit{Layers} dynamically 
activate/deactivate depending on the context currently executing, 
composing/withdrawing the behavioral variations associated to them 
with the base system behavior.   

Complex real-world self-adaptive systems consider several 
requirements. For instance, distribution leads to various contexts in 
different concurrent components. Correspondingly, if several 
components create different contexts, they may trigger behavioral 
changes in others. Thus, behavioral variation activation performed by 
asynchronous communication is desirable. Considering  
highly dynamic environments, performing behavioral changes without 
inconsistent/erroneous behavior is also a primary concern.

\textsf{ContextErlang}~\cite{salvaneschi2015contexterlang}, is a \ac{COP} language
based on context-aware agents. Agents have behavioral units that can be dynamically 
activated, \ie \emph{variations}. 
ContextErlang uses message passing as a variation activation mechanism. This mechanism 
leads to the asynchronous activation required for real-world applications. 

\textsf{ContextErlang} defines a core calculus semantics based on the 
actor concurrency model. The advantage of the language semantics is that the language 
capabilities can be easily replicated in other actor languages, for example in Scala. 
\textsf{ContextScala}~\cite{salvaneschi2015contexterlang} is a \ac{COP} language implemented on 
top of the \textsf{Akka} framework in Scala, based on \textsf{ContextErlang}.
In ContextScala variations are defined reusing the main modularity abstraction of Scala: classes. 
As an illustration, we show an example of variations definition from our running example with two variations of 
the move behavior to manage the direction of the movement, \scode{Up} 
(Lines 1-3), and \scode{Down} (Lines 4-6). 

\begin{lstlisting}[language=Scala, numbers=left, label={lst:variationExample}]
 class Up extends Variation[Up] {
   def moved(): Unit = println("Movement Up")
 }
 class Down extends Variation[Down] {
   def moved(): Unit = println("Movement Up")
 }
\end{lstlisting}

In \textsf{ContextScala} variation activation is managed by an 
instance of the ContextAgent (an \textsf{Akaa} actor), which sends
activation messages to the corresponding actor. The \scode{ContextAgent} manages 
context variations through the definition of the variation list, given to the 
\scode{SetActiveVariations} function as shown in the following  
snippet (Line 1). Upon a method call (the  \scode{moved} method in our 
example), the implementation corresponding to the activated variation 
is executed. In our example, Line 2 executes the definition of the 
\scode{Up} variation (activated in Line 1), and Line 4 Executes the 
definition of the 
\scode{Down} variation (activated in Line 3). 

\begin{lstlisting}[language=Scala, numbers=left, label={lst:variationActivation}]
 ContextAgent ! SetActiveVariations(List(Up()))
 ContextAgent ! moved()   //result: Movement Up
 ContextAgent ! SetActiveVariations(List(Down()))
 ContextAgent ! moved() //result: Movement Down
\end{lstlisting}

The use of layers and behavioral adaptations posits a modular 
way to define dynamic program adaptations, based on the surrounding execution context. 
The fine-grained partial behavior definitions in \ac{COP} enable us 
to adapt any system behavior at any moment in time, offering greater 
flexibility to realize self-healing systems without pre-define strategies.

\subsection{Reinforcement Learning Options}
\label{sec:QLearning}

\ac{RL}~\cite{sutton98} is a technique to learn optimal actions for 
specific environmental conditions by trial-and-error, based on 
interactions with the environment.
Q-learning is a common implementation of \ac{RL} agents that, for each time step, 
perceives the environment and maps it to a state {\color{purple}$s_i$} from 
its state space $S$. It then selects an action {\color{purple}$a_i$} from its action 
set $A$  and executes it. The agent receives a reward {\color{Bittersweet}$r_i$} from the 
environment when it transitions to the next state, based on which it 
updates the suitability of taking action {\color{purple}$a_i$} in state {\color{purple}$s_i$}. The 
agent's goal is to learn a policy (\ie the most suitable action for 
each state) to maximize the long-term cumulative reward. The learning 
rate {\color{NavyBlue} $\alpha$} determines the extent new experiences overwrite 
previously learned ones, and the discount factor {\color{RoyalBlue} $\gamma$}  determines 
how much the future rewards are discounted for agents to prioritize 
immediate actions but still be able to plan the best long term 
actions. At each timestep, Q-value of an action {\color{purple}$a_i$} taken in state {\color{purple}$s_i$} is updated using the Q-learning equation bellow. 

\vspace{0.7em}

\begin{equation*} \label{eq:QL}
\scriptstyle
    Q(s_{t+1},\, a_{t+1}) \leftarrow  
    {\tikzmarknode{qt}{\highlight{purple}{$Q(s_t, a_t)$}}} +
    {\tikzmarknode{alpha}{\highlight{NavyBlue}{$\alpha$}}}
    [ 
    {\tikzmarknode{r}{\highlight{Bittersweet}{$r_{t+1}$}}}  +  
    {\tikzmarknode{gamma}{\highlight{RoyalBlue}{$\gamma$}}}
    {\tikzmarknode{max}{\highlight{OliveGreen}{$\max\limits_a Q(s_{t+1},a)$}}} - 
    {\tikzmarknode{qt2}{\highlight{purple}{$Q(s_t, a_t)$}}} 
    ]
\end{equation*}

\begin{tikzpicture}[overlay,remember picture,>=stealth,nodes={align=left,inner ysep=1pt},<-]
    \path (qt.north) ++ (3.51,1.6em) node[anchor=south east,color=Mulberry!85] (ntext){\textsf{\footnotesize Q-value}};
    \draw [color=Mulberry](qt.north) |- ([xshift=0.8ex,color=Mulberry]ntext.south west);
    \path (qt2.north) ++ (-1.2,1.6em) node[anchor=south east,color=Mulberry!85] (qt2text){};
    \draw [color=Mulberry](qt2.north) |- ([xshift=-4.9ex,color=Mulberry]qt2text.south west);
    \path (alpha.north) ++ (-0.2,-2.8em) node[anchor=south east,color=NavyBlue!85] (atext){\textsf{\footnotesize learning rate}};
    \draw [color=NavyBlue](alpha.south) |- ([xshift=-9.3ex,color=NavyBlue]atext.south east);
    \path (r.north) ++ (-0.1,1.5em) node[anchor=north east,color=Bittersweet!85] (lijtext){\textsf{\footnotesize reward}};
    \draw [color=Bittersweet](r.north) |- ([xshift=-4.3ex,color=Bittersweet]lijtext.south east);
    \path (gamma.north) ++ (0.5,1.5em) node[anchor=north west,color=RoyalBlue!85] (gtext){\textsf{\footnotesize discount factor}};
    \draw [color=RoyalBlue](gamma.north) |- ([xshift=-2.9ex,color=RoyalBlue]gtext.south east);
    \path (max.north) ++ (-1.2,-3.6em) node[anchor=south west,color=xkcdHunterGreen!85] (lmaxtext){\textsf{\scriptsize Maximum Q-Value in the next state}};
    \draw [color=xkcdHunterGreen](max.south) |- ([xshift=-5ex,color=xkcdHunterGreen]lmaxtext.north);
\end{tikzpicture}

\vspace{0.15em}

Early environment interactions are focused on 
exploration, \ie actions are picked randomly and uniformly from the 
actions available in a given state. At the same time, after an agent 
has had a chance to learn the quality of actions, later interactions 
focus on exploitation, \ie mainly executing those actions 
known to lead to the highest long-term rewards. 

Whenever sequences of actions are frequently found together, \ac{RL} 
options package such sequences together as a means to optimize the 
agent execution by means of reducing the amount of decision points, 
for each of the actions to the entry point of the complete 
sequence~\cite{stolle02} or behavior history~\cite{girgin05}.  

Options are used in \ac{RL} to speed up learning or to minimize the 
periods of suboptimal performance during exploratory interaction with 
the environment. We consider \ac{RL}-based option 
techniques~\cite{sutton98intra}, to be suitable for learning and 
integrating sequences of recovery actions. Packaging options as \ac{COP} adaptations 
can be used to autonomously trigger sequences of recovery actions in self-healing systems 
in response to context changes (\ie failure detection), akin to the way actions in \ac{RL} are 
learned and taken in response to observed environment conditions.


\section{Learning to Heal in Unannounced Situations}
\label{sec:self-healing}

This section introduces our proposal for a self-healing framework designed to: 
\begin{enumerate*}[label=(\arabic*)]
\item detect faults flexibly using declarative monitor definitions,
\item learn and extract recovery strategies online from user/learned behavior to avoid  
pre-defined behavior, and
\item dynamically adapt behavior to heal failures, according to the systems execution context.
\end{enumerate*}
In the following, we present the complete self-healing process posit by our framework. 

\begin{figure}[htpb]
    \centering
    \includegraphics[width=0.45\textwidth]{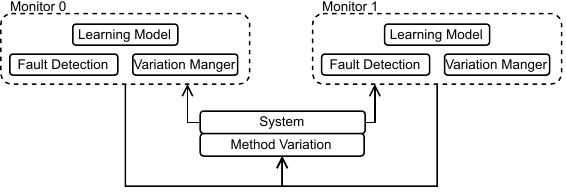}
    \caption{Monitor Architecture}
    \label{fig:MonitorArchitecture}
\end{figure}

The objective of our framework is to recover from 
unknown situations at run time. To achieve this, we position monitors 
at the heart of the solution, as in \fref{fig:MonitorArchitecture}. 

Monitors are used to handle fault detection. When first encountered, a point of failure is used to 
trigger learning recovery strategies process from actions executed by external agents 
(\eg users, automated agents). 
Having multiple possible strategies, the learning model the best suited 
corrective behavior from all learned options. Once the learning 
stage finishes, the variation manager generates the recovery strategy from action 
sequences, defined as a variation to execute at run time. The 
fault detection system, now triggers the appropriate learned 
variations as faults are detected.

The main contribution of the framework is that, opposed to existing 
self-healing systems, monitors are not constrained to specific application 
points (\ie hooks) where a known healing strategy can be applied. 
Rather, the declarative definition of monitors enables us to evaluate 
different system properties at any point during the execution. 
Furthermore, we are not restricted to applying pre-defined recovery 
strategies. Learning recovery strategies from combinations of 
actions in the base system behavior helps us recover from unannounced 
situations, a property unique to our approach.

\subsection{Fault Detection} 
\label{sec:error-detection}

Monitors are defined to observe the system execution. 
Monitors are instantiated as needed in the 
application, given their declarative definition, specifying the variables or 
functions to observe, rather than specifying fixed application 
points. \fref{fig:MonitorArchitecture} shows monitor's internal 
structure and interaction with the environment. Monitors are composed 
of: the fault detection system, the learning model, and the variation 
manager. 

Monitor instances require the specification of a set of relevant 
variables (\ie observable variables), and a predicate 
verifying such variables. Monitors verify the system behavior
based on REScala's observers; thus, the monitor verification
is triggered with every update to the variables under observation. 
For our running example, we define a monitor through the \scode{orFault} predicate in 
\fref{lst:or-predicate}, evaluating the ASCII 
code of the mouse position given by the \scode{x()} signal.

\begin{lstlisting}[language=Scala,
    frame=lines,
	label={lst:or-predicate},
	caption={Monitor predicate declaration}]
 val orFaults =  (x: Signal[Int]) => Signal[Boolean] { 
     (40 < x() and x() < 50) or 200 < x() 
   }
\end{lstlisting}

Moreover, monitors can refine the predicates dynamically, for example, if the acceptable conditions 
of the behavior change at run time. This 
is achieved by defining new predicates on top of the existing ones to 
account for the new conditions. New predicates can overwrite (as in 
\fref{lst:prime-predicate}) or complement the behavior of  
existing predicates. In our example, the \scode{primeFaults} predicate is 
used to overwrite the \scode{orFaults} predicate. 

\begin{lstlisting}[language=Scala,
    frame=lines,
	label={lst:prime-predicate},
	caption={Primality check predicate declaration}]
 val primeFaults = (x: Signal[Int])=>Signal[Boolean]{ 
     isPrime(x()) 
  }
\end{lstlisting}

\subsection{Recovery Strategy Generation} 
\label{sec:resolution-generation}

The second step in generating recovery strategies involves learning corrective action sequences 
to transition from a fault state to a valid state. Based on \ac{RL}, learning first undergoes a training 
phase, in which the system executes (atomic) system actions for a set number of steps, dependent on the 
state and action space of the system. During each step, whenever a monitor detects a
fault state, it initiates the learning process.
The learning process requires to explore actions, and based on an obtained reward decide if the 
effectively correct the system behavior. corrective behavior. Once the the reward on actions taken 
converges, we exploit these learned actions
as a recovery strategy for the detected fault state. Following this, our framework is
designed for continuous learning, where new recovery actions can be learned even after a
recovery strategy has been generated already.

In particular, we use a Q-learning agent to learn atomic action sequences --that is,
behavior (\eg functions, methods) defined in the base system. The monitors continuously
capture the system state for each event, evaluating their corresponding predicates. Once
failure states are detected, as events not satisfying the monitor's predicate, we use the
learning model to record the set of atomic actions used to bring the system back to a
valid state. To drive the learning process, each atomic action must be associated with a
reward, to maximize the attainable reward.


After the learning process concludes, the learning model generates a map that encapsulates
the fault states and a list of the learned recovery actions, containing those atomic actions with the 
highest reward from those states, as illustrated in \fref{lst:learning}.
These highest reward atomic actions, depending on the context, can be used to construct
complex healing strategies by composing such atomic actions with the highest rewards.
Whenever the monitor detects a fault of an already learned state in the map, the system
now will automatically execute the healing strategy, until it reach a valid state.

\begin{lstlisting}[language=Scala,
    frame=lines,
	label={lst:learning},
	caption={Learning atomic actions}]
val DetectionObservable = observable observe {     
  State => if (learningFlag) {
    step(State.oldPosition, State.newPosition)
  } else if (tableVariationCreationFlag) {
    agentData.Q.foreach { 
      case (key, value) => 
      StateOfVariations.updateState(key, updateCurrentVariationState(key, value, List())) 
    }
    tableVariationCreationFlag = false
  }
}
\end{lstlisting}

Note that the effectiveness of the generated healing strategies is contingent on both the
expressiveness of the predicates and the exploration of possible states during the
learning stage. If the monitor only learns from local states, there won't be any generated
healing strategies for states not explored during learning. Moreover, if the predicate's
expressiveness falls short in identifying desired fault states, the learning stage may
overlook them, leading to incorrect reactions.

\subsection{Healing From Error States} 
\label{sec:healing-strategy}

The final stage in our process consists on autonomously applying the 
healing strategy at any given application point of the system 
(definition or execution) rather than relying on predefined actions 
for particular program points. 

Once a fault state is detected, and the association between the state 
and its sequence of atomic actions learned, the system should be 
capable to enact the sequence at the fault state, rather than falling 
into it. The dynamic composition and withdrawal of behavior of 
\ac{COP} is used for this purpose. 
That is, whenever a fault state is reached and flagged by the 
monitor's predicate, the system should automatically execute the 
learned sequence of atomic actions to recover from said failure.

To do this, whenever a predicate is triggered, if its state is defined 
in the reaction map, the monitor interrupts system's base execution 
and proceeds to execute the sequence of recovery actions, as shown in 
\fref{lst:cop}. Once the reaction is finished, the expected behavior 
control execution is returned to the element. 

\begin{lstlisting}[language=Scala,
    frame=lines,
	label={lst:cop},
	caption={Automated application of recovery strategies}]
def takeAction(ctx: Boolean, e: MouseMoved): Unit = 
if(ctx) {
  val agent = AgentLocation(e.point.x, e.point.y)
  val variationList: List[QMove] = StateOfVariations.getState.getOrElse(agent, List())
  if (variationList.nonEmpty) {
    val castedVariationList = CastingDefinition.QMoveListToDefinitionList(variationList)
    actor ! SetActiveVariations(castedVariationList)
    actor ! moved(e.point.x, e.point.y)
  }
} else 
currentObservable.mouse.mouseMovedE.fire(e.point)
\end{lstlisting}

Given that the recovery strategy corresponds to a list of atomic 
actions, the execution of the strategy consist of setting the list of 
variation to the variation manager and then calling the system 
behavior, which now corresponds to that of the first variation in the 
list; as each variation finishes, it calls upon the next variation, 
leading to the execution of all the variations in the sequence. 
Finally, after the execution of the healing strategy the system 
reverts to the base system behavior.

In our running example, getting to state $(17, 28)$ will trigger the 
monitor invalidating the \scode{orFaults} predicate in 
\fref{lst:or-predicate}, as $10 < 17 + 28 < 50$. During exploration, our framework learns to recover from this state by generating sequences of atomic actions (\ie mouse movements).


\section{Validation}
\label{sec:validation}

This section shows the feasibility and effectiveness of our framework for the self-healing 
of reactive programs, without having to specify program hooks or strategies beforehand. The 
validation consists of two different 
applications from different domains and knowledge areas.
First, we present our running example of a prototypical application 
for reactive systems, taken from the REScala 
exemplars.\footnote{https://github.com/rescala-lang/REScala/tree/master/Code/Examples} 
This application demonstrates the effectiveness of our 
approach to heal systems in the reactive systems application domain.
Second, we use the \deltaiot self-healing exemplar to evaluate the 
effectiveness of our approach in generating appropriate healing 
strategies (\ie in line with predefined strategies) without previous specification, to manage unforeseen 
situations.

\subsection{ASCII Mapping Tessellation}
\label{sec:proof-of-concept}

We validate the feasibility of our framework using the mouse movement 
tracking application. In particular, we demonstrate it is possible to: 
\begin{enumerate*}[label=(\arabic*)]
\item define monitors for different system behavior/states,
\item learn strategies that recover from failure states autonomously.  
\end{enumerate*}

\subsubsection{Description}

The mouse position is represented as a signal that updates whenever the mouse moves on a
$100 \times 100$ GUI, as illustrated in \fref{fig:running}. Using the mouse position signal, we 
define two evaluation scenarios according to different distributions of faulty states.

Fault states are defined as tessellation patterns based on the number given by the mouse's
position in the GUI. Tessellations are defined using the predicates to detect failures.
The predicates are formulated based
on REScala's built-in functionality and user behavior. Our scenarios are based on two
predicates, \scode{orFaults} (\fref{lst:or-predicate}), and \scode{primeFaults}
(\fref{lst:prime-predicate}), using the logical \scode{or} operator and a predicate to
determine primality, respectively.  The \scode{sum} signal converts the $(x, y)$
coordinates into an integer that is input to the two predicates.

\begin{scala}
 val sum: default.Signal[Int] = Signal[Int] {
    (mouseX() + mouseY()) % 255
 }
\end{scala}

\begin{figure}[hptb]
     \centering
     \begin{subfigure}[t]{0.43\columnwidth}
         \centering
         \includegraphics[]{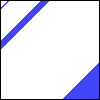}
         \caption{Tessellation using the \scode{orPredicate} with three specific fault regions}
         \label{fig:tessellation-predicateTwo}
     \end{subfigure}     
     ~
     \begin{subfigure}[t]{0.43\columnwidth}
         \centering
         \includegraphics[]{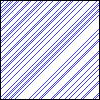}
         \caption{Tessellation using the \scode{primePredicate} with multiple fault regions spread through the GUI}
         \label{fig:tessellation-predicateOne}
     \end{subfigure}
     \hfill
    \caption{Errors arise in the blue areas according to the two predicates}
    \label{fig:tessellations}
\end{figure}

The tessellations generated by the predicates are respectively shown in
\fref{fig:tessellations}, where the blue region represent the failure GUI states.
Mouse movements trigger the signals. Failure detection is then a reaction associated
with the monitor. 
Monitors actively observe mouse movements to detect faults (blue regions) using the defined
predicates (\scode{orFaults} or \scode{primeFaults} in our case), then start corrective actions 
to move the mouse to a white region.
The tessellations in the two scenarios generate different fault distributions in the GUI states, as 
a case where no predefined corrective action fits both scenarios. 
In \fref{fig:tessellation-predicateOne}, failure states
are more uniformly distributed than in \fref{fig:tessellation-predicateTwo} where faults
are concentrated in three regions. 

As discussed in \fref{sec:QLearning}, a predefined set of
possible actions must be established. In this context, the actions correspond to mouse
movements: \scode{Up}, \scode{Down}, \scode{Left}, and \scode{Right}. These actions
are the atomic actions available to the system to correct the mouse position between
blue and white regions. Additionally, we define a reward function that
assigns a value of 1 when the predicate is true (indicating the mouse
is in a valid position) and 0 when false. Therefore, each mouse position $(x, y)$ maps to a 
collection of rewards, for each of the possible actions in that position, the Q-value map.

The execution process of the system is as follows: 
\begin{enumerate*}[label=(\arabic*)]
\item the mouse moves through the GUI in a random path ensuring that all positions have the same probability of being visited.
\item During the initial part of the process, actions moving from failure states to valid states are 
recorded with a positive reward.
\item Once learning concludes, the Q-value map encapsulates optimal atomic actions for each GUI state.
\item We systematically concatenate individual optimal atomic actions in failure states, until a valid state is reached, effectively generating a healing strategy for the initial failure state. 
\item Finally, whenever a failure state is detected again, the learned corrective actions sequences are applied autonomously, taking the mouse to a valid state, and continue with its movement (step 1).
\end{enumerate*} 

Note that loops may appear in the construction of healing strategies. The process 
must account for loops and transitive relations within the path, to improve the process. 
state. This formalized approach ensures a methodical exploration of potential action
sequences and their efficacy in guiding the system towards valid states.

\subsubsection{Experimental Design}
\label{sec:expiremental-desing}

We consider two tessellations to evaluate the effective generation of strategies in different 
scenarios. The scenarios correspond to two predicates: 
(\scode{orPredicate}, \scode{primePredicate}), used by the monitors. 
In both scenarios
an external automatic agent moves the mouse uniformly through the whole GUI for both experiments 
assuring all states are visited, running for 100000 steps, a step is a movement of the mouse. 
Each experiment runs for five iterations.

The first experiment counts the number of times the monitor detects a failure, to compare the 
effectiveness in detecting failures against the theoretical value. Once learning finishes, the second 
experiment takes place. The agent moves the mouse on failure regions to test how many healing 
strategies generate reactions leading to valid states. 

\subsubsection{Results}
\label{subsec:results}

\fref{tab:results} shows the result of the experiments where: 
\begin{description}[leftmargin=0.1cm]
  \item[Faults detected] shows the number of failure states that the monitor detected through the learning phase. That is, out of the 100000  executed steps how many triggered the monitor's predicate. 
  \item[Fault proportion] shows the proportion of detected faults to the number of steps (100000). 
  \item[Correct strategies] shows the number of healing strategies that lead to a correct state, once the learning phase ends. To get this value, for every fault state in the tessellation, we check whether or not the associated healing strategy leads to correct state.
  \item[Healing effectiveness] shows the proportion of correct strategies to the
  number of states to heal (\ie how well the system heals after the learning phase).
  The number of states to heal each tessellation is
  1174 for \scode{primePredicate}, and 2090 for \scode{orPredicate}. 
\end{description} 

\begin{table*}[htp]
  \centering
  \caption {Experimental iterations results}
  \begin{tabular}{c c c c c c || c c c c c } 
  \multicolumn{1}{c}{}& \multicolumn{5}{c}{\textbf{orPredicate}}& \multicolumn{5}{c}{\textbf{primePredicate}} 
  \\
  \toprule
  \multicolumn{1}{c}{}& 
  \multicolumn{1}{c}{\textbf{1}}& 
  \multicolumn{1}{c}{\textbf{2}}& 
  \multicolumn{1}{c}{\textbf{3}}&
  \multicolumn{1}{c}{\textbf{4}}&
  \multicolumn{1}{c}{\textbf{5}}&
  \multicolumn{1}{c}{\textbf{1}}&
  \multicolumn{1}{c}{\textbf{2}}& 
  \multicolumn{1}{c}{\textbf{3}}&
  \multicolumn{1}{c}{\textbf{4}}&
  \multicolumn{1}{c}{\textbf{5}}
\\
\toprule
 & \\[\dimexpr-\normalbaselineskip+5pt]
 \textbf{Faults detected}           & 18122  & 18390  & 18127  & 18404  & 18048  & 20786  & 20652  & 20594  & 20712  & 20805   \\ 
 \textbf{Faults proportion}         & 18.1\% & 18.3\% & 18.1\% & 18.4\% & 18.0\% & 20.7\% & 20.6\% & 20.4\% & 20.7\% & 20.8\%  \\
 \textbf{Correct strategies}        & 649    & 797    & 309    & 686    & 843    & 1937   & 1786   & 1677   & 1808   & 1842    \\ 
 \textbf{Healing effectiveness}     & 55\% & 67.8\% & 23.6\%   & 58.4\% & 71.8\%   & 92.6\% & 85.4\% & 80.2\% & 86.5\% & 88.1\%  \\ 
 & \\[\dimexpr-\normalbaselineskip+2pt]
 \bottomrule
  \end{tabular}

  \label{tab:results} 
\end{table*}

The first two rows in \fref{tab:results} measure the monitor 
failure detection. For the \scode{primePredicate} tessellation $13.3\%$ of states are failures states, 
and for the \scode{orPredicate} $26.4\%$ are failure states. 

The last two rows in \fref{tab:results} measure monitor failure and healing strategies effectiveness. 
For instance, the mean value for correct strategies in the tessellation of the \scode{primePredicate} 
has a high precision, with a low standard deviation. This leads us to conclude that the monitor is 
precise and accurate when generating and executing healing strategies. 

At first glance, the result for the \scode{orPredicate} tessellation seems contradictory. 
For example, iteration 3 shows a healing effectiveness of $23\%$, with a mean value of $55.7\%$ 
for the five iterations, and a standard deviation of 16.98\%. 
A possible reason for these values come from the density of failure states in each scenario.
\fref{fig:tessellations} shows the GUI covering with tessellations using a different distribution of 
failure states. The \scode{orPredicate} tessellation has three specific regions containing all 
failure states, in failure-dense regions learning corrective strategies will take longer. In contrast, the 
\scode{primePredicate} tessellation has a better propagation of the reward due to the many correct 
regions surrounding the failure regions. Therefore, the amount of learning steps for the systems 
should be defined by users, based on the 
system complexity and interaction with failure states observed by the monitors.  

%

\subsection{\deltaiot Exemplar}
\label{sec:exemplar}

The \deltaiot exemplar is a multi-hop LORA communication network composed of 25 
\ac{IOT} motes deployed in different physical locations~\cite{iftikhar17}. \deltaiot is 
used as an exemplar to evaluate different self-adaptation strategies to manage motes' 
trade-off between battery consumption and packet delivery.\footnote{undisclosed repo url}

\subsubsection{Experimental Design}

The \deltaiot project\footnote{\url{https://people.cs.kuleuven.be/~danny.weyns/software/DeltaIoT}} 
provides a simulation environment of a LORA network, with simulated behavior for the battery 
consumption and package handling of the different motes, as a subset of the real-world 
implementation. In the simulation, there are 96 execution steps, in each step every $(Mote,Link)$ 
combination is traversed.
During each step, the system uses predefined adaptations to improve the \scode{PacketLoss} and 
\scode{EnergyConsumption} metrics for the motes.

To apply our approach in the \deltaiot exemplar, we remove the prescribed adaptations and define the 
execution steps of the simulation using reactive signals. The main signal represents the execution of 
the simulation, in which an event is a $(Mote,Link)$ combination. Building upon signals, we establish 
the monitoring system. Monitors leverage the activation strategy embedded in the simulation. In 
particular, we create a predicate, \scode{analyzeLinkSettings}, measuring links' state with respect to 
the two metrics. When the predicate is satisfied, it triggers the deployment of the variations defining 
the learned healing strategies. The monitor, guided by a learning model, 
efficiently manages healing adaptation strategies, ensuring effective surveillance of the simulation, 
as described in \fref{sec:self-healing}.
While we reimplemented the use of monitors and simulation events, our experiment mirrors 
the original simulation. The Scala extension to the project is modularly independent from the original 
\deltaiot implementation.

In the learning phase, every time a monitor's predicate is triggered, the user chooses the corrective 
actions for the energy use or path in the network configuration. We run \deltaiot for 96 execution steps, 
as training of our \ac{RL} model (the original execution steps of the simulation). Once variations are 
learned, we run the simulation for 54 additional steps, that do not depend on the specific learning 
hooks. During these steps, the learning model leads the healing strategies to execute in the system 
from the learned behavior.
In this experiment, managing mote energy and packet loss at par with the prescribed adaptations 
from \deltaiot is consider a success for our solution.

\subsubsection{Results}

\begin{figure}[htpb]
    \centering
    \includegraphics[width=0.9\columnwidth]{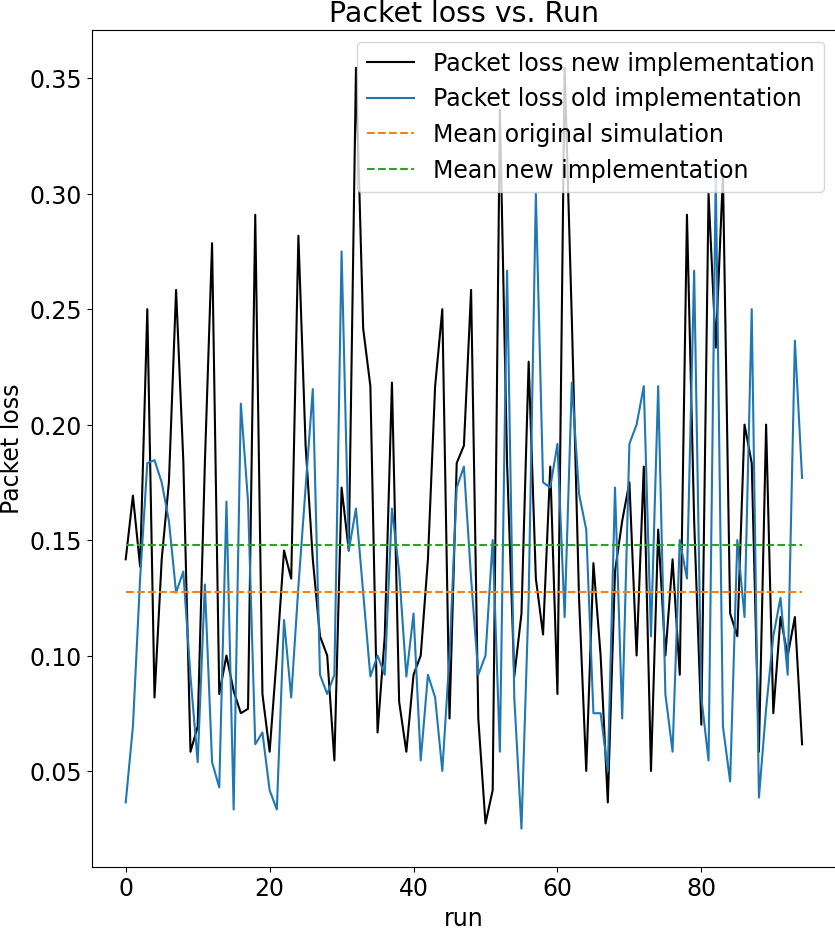}
    \caption{Packet loss values after both simulation execution}
    \label{fig:packetloss}
\end{figure}

\begin{figure}[hptb]
    \centering
    \includegraphics[width=0.9\columnwidth]{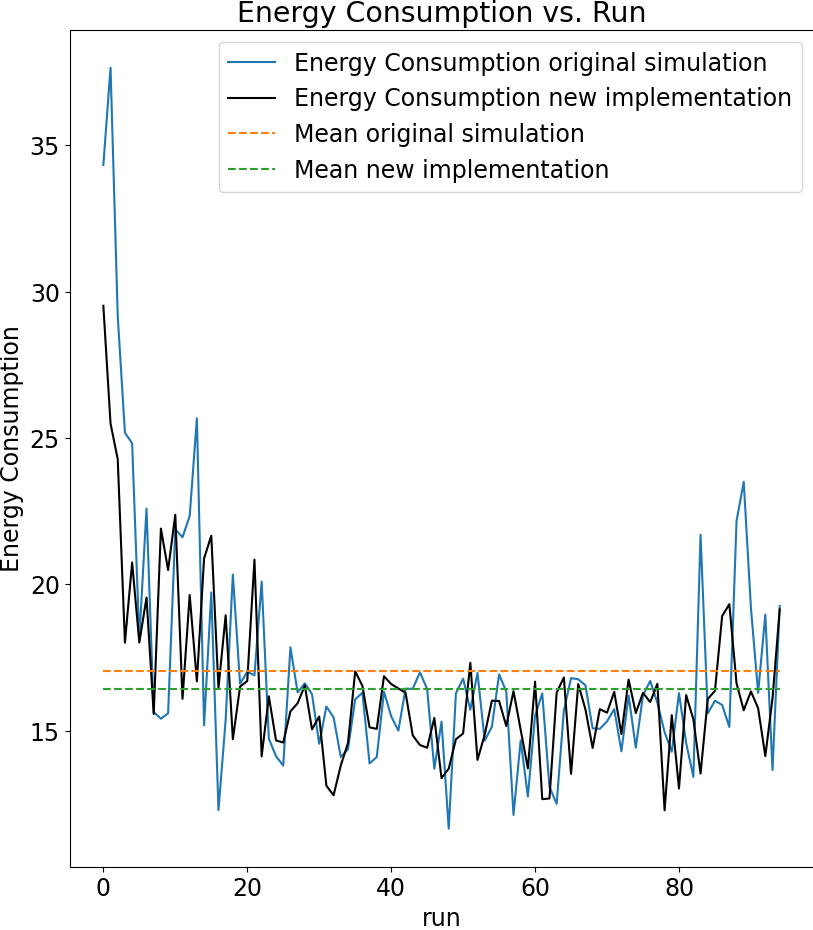}
    \caption{Energy consumption values after both simulation execution}
    \label{fig:powerconsumption}
\end{figure}

\begin{figure}[hptb]
    \centering
    \includegraphics[width=0.9\columnwidth]{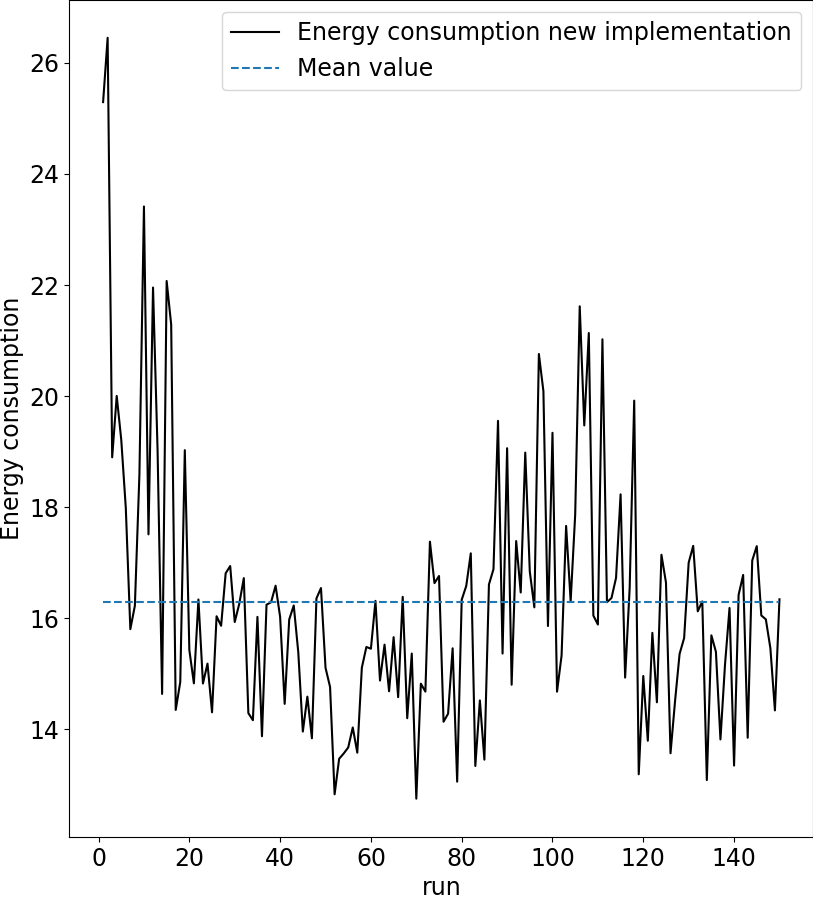}
    \caption{Power consumption values for new implementation long run}
    \label{fig:powerconsumptionlong}
\end{figure}

The results show that the strategies learned from the application usage, coincide with the prescribed
adaptations in the original \deltaiot simulation~\cite{iftikhar17}. Figures \ref{fig:packetloss} 
and \ref{fig:powerconsumption} display the results for the packet loss and power consumption 
adaptation strategies, respectively. The behavior of our approach with respect to packet loss 
\fref{fig:packetloss} maps to that of the original simulation, the mean packet loss in our solution is 
slightly higher than that in the original implementation. This might be due to the higher rate of lost 
packets in the first steps of the simulation, while the algorithm is still learning. However, towards the 
last steps, once learned variations are in place, we observe a far lower rate of lost packets. 
The energy consumption case, \fref{fig:powerconsumption} presents a similar situation, with a slight 
improvement seen using our approach. The results allow us to conclude that learning the adaptation 
strategies is as effective as using the predefined adaptations in reaching the system's objective, 
with the added value of not having to provide the pre-defined adaptations or the hooks to monitor and 
apply them. Additionally, \fref{fig:powerconsumptionlong} shows the life-long learning and continuous 
behavior of our proposal. Here, the learned strategies continue working without the need for any 
particular self-healing strategy, and the system can continue a learning more strategies.

\fref{tab:resultsDeltaIoT} shows the behavior of both implementations from the perspective of detected 
faults and the corrective actions applied. With these results we can assure that the system is behaving 
as it is supposed to, without the need for particular predefined hooks in the application.

\begin{table}
  \centering
  \caption {Event detection \deltaiot vs. our framework}
  \begin{tabular}{c c c} 
  \multicolumn{1}{c}{}   & \textbf{Proposed solution} & \textbf{DeltaIoT} \\
\toprule
 \textbf{Total events}             & 603   & 603  \\ 
 \textbf{Correct strategies}       & 411    & 420  \\ 
 \bottomrule
 \end{tabular}

  \label{tab:resultsDeltaIoT} 
\end{table}


\section{Related Work}
\label{sec:related-work}

This section presents a comparison between related self-healing  
approaches, existing \ac{COP} solutions using \ac{RL}, and our 
proposed approach.

\subsection{Self Healing}
\label{sec:self-heling}

Early work in this field focused on a self-stabilizing system where the system reaches a
legitimate state in a finite number of steps regardless of its initial
state~\cite{10.1145/361179.361202}. Following similar definitions, several approaches have
been proposed to define self-healing systems. For example, a classification by research
area is presented by~\citet{Psaier2010}. Here the authors identify various implementations
of self-healing systems according to their application domain: embedded systems, operating
systems, reflective middleware, and web services, among others.
~\citet{Koopman2003ElementsOT} presents a set of general concepts that all self-healing
systems have in common, regardless of their application domain. In this sense, the system
must have a fault model with complete knowledge of the faults expected to self-heal,
having some information regarding fault duration, fault manifestation, and fault source.
Besides, the system must be complex enough to respond to the fault, meaning that the
proper mechanism for fault detection, response, and recovery must be considered.
Furthermore, self-healing approaches must evaluate the system completeness. 
it must account for how the system behaves over time and handle the various changes in the
system architecture that may occur at run time and architectural completeness. Finally,
~\citet{Koopman2003ElementsOT} proposes that the design context shapes in a particular way
the self-healing capabilities. 

Concerning the architectural completeness, before any self-healing system 
can occur, a significant infrastructure must be put in place to support fault detection and 
repair~\cite{dashofy2002towards}. 
Specifically, the system must be built using a framework that provides: run-time adaptation, a language 
to express the repair plan, and an agent to execute the repair.  

Usually, self-healing systems are tightly integrated with the application. Granting the
systems the ability to deal with failures at detection. Nonetheless, externalization
mechanisms are proposed \cite{10.1145/582128.582134}, where the self-healing system is
untangled from the base application using an architectural model approach to design the
monitoring, problem detection, and repair; i.e., the self-healing system works on top of
the base application understanding what the running system is doing in high-level terms.
Following this idea, \citet{5395114} proposes that these systems are suitable for
continuous learning. The system could learn from its run-time behavior to improve its
capabilities. Specifically, the case study of this paper presents a client-server
application in which the self-healing system is based on a PRNN (Pipelined Recurrent
Neural Network) where the parameters are the server status and the bandwidth level.
Another example of learning capacity for self-healing systems
\cite{10.1007/978-3-642-10665-1_5} uses a Multivariate Diagram and a Naive Bayes
Classifier to determine severity levels and infer possible consequences. In this approach,
the model continuously changes its parameters based on the healing process. Reinforcement
Learning \ac{RL} has also been used for this application. For instance,
\citet{schneider2015survey} shows self-healing systems' capabilities in the area of LTE,
implemented using a RL scheme. Similarly, \citet{5671622} proposes a self-optimization
solution of coverage and capacity in LTE networks using fuzzy RL while operating complete
autonomous in a fully distributed environment. Additional research efforts exist to define
self-healing systems over networks~\cite{6214334,ANGSKUN2010479,265711}, using similar
techniques to the ones described before.

Moreover~\citet{zhao2017reinforcement} propose a reinforcement learning-based framework
for the generation and evolution of adaptation rules in software-intensive systems. The
framework involves an ofﬂine learning phase, similarly to our proposal, to generate
adaptation rules for various goal settings and an online adaptation phase to use and
dynamically evolve these rules. It addresses the limitations of traditional rule-based
adaptation, such as guaranteeing optimal adaptation and supporting rule evolution to cope
with non-stationary environments and changing user goals at runtime. 
Similar to our approach of learning actions at run time to heal the system through adaptations, ~\citet{zhang2021meta}  introduce self-learning adaptive systems (SLAS)
and underscore the significance of acquiring high-performance adaptation policies for
dynamic scenarios. SLAS proposes a new method called meta reinforcement learning adaptive
planning (MeRAP) for the online adaptation of SLAS, which separates concerns related to
adaptation policy, models environment-system dynamics, and employs a meta reinforcement
learning algorithm for offline training and online adaptation.

\subsection{Combing COP and Learning Approaches}
\label{sec:COP-Learning}
Our self-healing approach incorporates the Q-learning reinforcement learning approach, enabling the system to define healing decisions automatically. Then, when the system detects a failure state, this healing strategy is executed automatically using context-aware variations of the defined methods.

Similarly, \citet{cardozo2018generating} propose to have no predefined adaptation using a 
proof-of-concept to illustrate this idea while showing the challenges of implementing such systems. 
For instance, the systems should be capable of integrating new elements, data sources, atomic actions, 
and goals at run time. Building on that, Auto-COP~\cite{2103-06757} uses \ac{RL} to build action 
sequences based on previous instances of the system execution.

As shown in \fref{sec:ContextScala}, \ac{COP} is used to generate dynamic behavioral responses to the system execution context; nevertheless, due to different simultaneous sensed situations, many adaptations could be applicable, which leads to conflict in systems' execution. An automated conflict resolution mechanism ~\cite{cardozo2017peace}, where W-Learning (an \ac{RL} algorithm) is used to capture the relationships between simultaneously proposed adaptations over time, updating their appropriateness as the system progresses.

Based on the revision of the current state of the art, it is possible to conclude that:
\begin{itemize}
    \item All self-healing systems must present the same general properties.
    \item Learning exists to try to manage self-healing systems, but there is no consensus, though RL looks promising.
    \item The application domain seems to be restricted for distributed systems.
\end{itemize}

We note that all current 
self-healing systems must have previous knowledge of the possible fault 
states in centralized systems --that is, only deal with known unknowns. Finally, to the authors' best knowledge, there is no work done on 
self-healing systems on centralized systems or reactive applications.



\section{Conclusion and Future Work}
\label{sec:conclusion}

There are two principal concerns when designing a self-healing system: detection and healing. 
The main challenges behind these concerns are the detection of fault states, the definition of healing 
strategies, and the definition of hooks for adaptation to introduce correct(ive) behavior. This paper 
proposes a framework to deal with the complexity of real word self-healing reactive applications. The 
framework proposes \emph{monitors} which are component abstractions defined to deal whit the 
challenges presented above. Monitors encapsulate the complexity of fault detection and adaptation 
definition and introduction in self-healing systems. Monitors enable developers to customize detection 
strategies and modularly observe the system at different granularity levels throughout is evolution. 
With this, the design and implementation of self-healing systems becomes more straightforward. 
 
The two key features of the proposed framework are: 
\begin{enumerate*}[label=(\arabic*)]
    \item A self-healing process driven by flexible \emph{monitors} defining fault detection at any granularity level, and applicable at any program location or moment in execution without predefined hooks.
    \item No need for predefined healing strategies. Our framework is able to learn healing strategies from the system's execution, generating sequences of actions that take the system from fault state to valid ones.
\end{enumerate*} 

The decrease in complexity and difficulty of detecting and handling failures in massive reactive 
systems, like streaming companies, makes this proposal appealing since monitors could be placed 
inside any existing services, automating the on-line recovery from errors, while avoiding the process 
of defining and managing complex healing strategies. We demonstrate the feasibility of our approach 
with a prototypical reactive application for tracking mouse movements in which several signals are 
generated using the mouse movement information. We introduce monitors to dictate fault and valid 
states of the system, automatically generating actions sequences to move from fault states to valid 
ones. With this, we expand the applicability of self-healing systems to the domain of reactive systems. 
We further demonstrate the effectiveness of our framework by applying it to the \deltaiot self-adaptation 
exemplar. In this application we demonstrate how the system is able to generate healing strategies that 
behave at par with those originally prescribed in the system. Furthermore, our model can continue 
learning to generate healing strategies for situations not previously conceived. 

As avenues for future work, we propose defining communication strategies between monitors
to enhance global healing strategies using local strategies, all without the need for
centralized orchestration. In alignment with this objective, we suggest designing
distributed reactive monitors as a technique for generating comprehensive global healing
strategies.

Additionally the learning phase is a time and resource intensive process. However, in this
work, our primary focus is not on addressing these resource concerns but rather on
highlighting the capability to generate effective healing strategies without the need 
of predefined hooks. For future work, we propose validating the efficiency of this
solution against predefined self-healing strategies, taking into account the
resource implications of the training phase as a determining factor.

\section*{Acknowledgments}

This work was sponsored, in part, by the Science Foundation Ireland (SFI) under Grant No. 18/CRT/6223 (Centre for Research Training in Artificial Intelligence), and SFI Frontiers for the Future project Clearway Grant No. 21/FFP-A/8957.

\printbibliography

%

\end{document}

\endinput